\documentclass[journal,twoside,web]{ieeecolor}
\usepackage{generic}
\usepackage{cite}
\usepackage{amsmath,amssymb,amsfonts}
\usepackage{algorithm}
\usepackage{algpseudocode}
\usepackage{graphicx}
\usepackage{textcomp}
\usepackage{lipsum}
\usepackage{bm}

\begin{document}
\title{An unsupervised segmentation of vocal breath sounds}

\maketitle

\begin{abstract}
Breathing is an essential part of human survival, which carries information about a person's physiological and psychological state. Generally, breath boundaries are marked by experts before using for any task. An unsupervised algorithm for breath boundary detection has been proposed for breath sounds recorded at the mouth also referred as vocal breath sounds (VBS) in this work. Breath sounds recorded at the mouth are used in this work because they are easy and contactless to record than tracheal breath sounds and lung breath sounds. The periodic nature of breath signal energy is used to segment the breath boundaries. Dynamic programming with the prior information of the number of breath phases($P$) and breath phase duration($d$) is used to find the boundaries. In this work, 367 breath boundaries from 60 subjects (31 healthy, 29 patients) having 307 breaths are predicted. With the proposed method, M ($89\%$), I ($13\%$), D ($11\%$) and S ($79\%$) is found. The proposed method shows better performance than the baselines used in this work. Even the classification performance between asthmatic and healthy subjects using estimated boundaries by the proposed method is comparable with the ground truth boundaries.


\end{abstract}

\begin{IEEEkeywords}
Dynamic programming, Breath sound, Asthma, Segmentation, Vocal sounds
\end{IEEEkeywords}

\section{Introduction}
Breathing is an irreplaceable process of human survival. Irregular breathing rate is one of the vital signs to indicate underlying poor psychological states like stress, anxiety as well as physiological conditions like cardiac arrest, asthma, COPD, etc. \cite{ritz2003behavioral}. To determine lung health, breath sound analysis is an emerging technique among physicians and researchers. Breath sounds can be recorded from the chest, trachea, nose, and mouth. Breath sound recorded at the chest is also referred to as lung sounds \cite{zimmerman2019lung}. Lung sounds show different acoustics properties between healthy and patients suffering from lung pathologies like pneumonia, asthma, etc. Adventitious lung sound are clinically characterized by their duration in respiratory cycles \cite{sarkar2015auscultation}.
Forgacs et al. \cite{forgacs1971breath} have noticed similar trends of different acoustic properties between healthy subjects and patients in breath sounds recorded at the mouth. 

One of our research interest is vocal sound based asthma monitoring and diagnosis. First work in this thread was the classification between asthmatic and healthy subjects using sustained phonations, cough and breath sounds recorded at the mouth \cite{yadav2018comparison}. Yadav et al. \cite{yadav2018comparison} have shown that among all the sounds breath sound performed the best. In the work by Yadav et al. \cite{yadav2018comparison} breath sounds manualy marked by the visual inspection of spectrogram and listening, which is a very time-consuming task.  Another work by the Yadav et al. \cite{yadav2021role} have shown that the classification performance between asthmatic and healthy subjects is better with prior knowlegde of breath boundaries as compared to randomly picked segments of breath signal from continuous breath cycles for the similar task.


Various methods have been reported for the segmentation of breath sounds recorded at the chest and trachea but the breath segmentation of breath sounds recorded at mouth is least explored. Palaniappan et al. \cite{palaniappan2017adaptive} have used normalized average power spectral density for breath phase detection (phases referred to inhale and exhale part of breath) as well as segmentation by using adaptive neuro-fuzzy inference systems. Aras et al. \cite{aras2018automatic} have used the dynamic time-warping algorithm to determine the boundaries of the respiratory cycles based on the similarity of the power spectral envelope between breath signals. The quasi-periodic nature of single-channel lung sounds' short-term energy is exploited to configure bandpass filters to estimate the respiratory phase boundaries \cite{yildirim2008automated}. Cam et al. \cite{le2008acoustical} have used the triplet Markov chain in a Bayesian framework to determine the breath boundaries and detect the phases. 
Feng et al. \cite{feng2008application} have proposed segmentation of tracheal breath sounds and their phases using Walsh basis functions. Jin et al. \cite{jin2009acoustical} have proposed respiratory phase segmentation in single-channel tracheal sound through the multi-population genetic algorithm with sample entropy as an evaluation metric. Yap et al. \cite{yap2001respiratory} have suggested a time-domain technique to detect respiratory onset using variance fractal dimension in tracheal breath sounds.
Yahya et al. \cite{yahya2014automatic} have used the nasal airflow to find the breath boundaries using a voice activity detector to detect the silence and breath phases. Detected voiced regions' peak values are used to detect the inhale and exhale phase by using a support vector machine.

In this work, we have proposed an algorithm for the segmentation of breath sounds recorded at the mouth (referred as vocal breath sounds (VBS)). Unlike breath sounds at the chest and trachea, the microphone can easily record breathing sounds at the mouth with minimum effort and no physical contact with the patient. Vocal breath sound data, we used in this work, is recorded in the hospital's natural noisy environment, making the proposed method suitable for real-time deployment. The database consists of healthy and asthmatic patients' breath samples; hence, the proposed method also considers the variability of normal and abnormal breathing rates. As most of the breath signal information is present in a frequency range up to 2kHz, all breath samples are low pass filtered to 2kHz. Nearly periodic nature of the VBS energy has been exploited to find out the boundaries of VBS and its phases ('inhale' and 'exhale'). VBS boundaries has been estimated using dynamic programming based approach with prior breathing duration and number of breaths information. A method to estimate breath duration and number of breaths have also been proposed in this work. As there is no existing work related to VBS segmentation, we used unsupervised phoneme segmentation \cite{dusan2006relation} and lung sounds segmentation techniques \cite{aras2018automatic} as baselines. Evaluation metric has been used to quantify, matched, missing, inserted, and deleted boundaries as given in the work by Ghosh et al. \cite{ghosh2007speech}. These metrics tells are calculated based on the distance between predicted and ground truth boundaries (helps in quantify matched boundaries), and difference between number of true boundaries and predicted boundaries (used to quantify missing, inserted and deleted boundaries). On the matched case, another evaluation metric overlap rate \cite{paulo2004automatic} has been used to determine the overlap between predicted and true boundaries. From the proposed method, we have found 89\% boundary matched out of them 79\% segment match with overlap rate (mean(standard deviation)) of 88($\pm 13$)\%). The proposed method outperforms both the baseline methods. Performance of the proposed method between the healthy and asthmatic group is also reported to analyze the algorithm's robustness to variable breathing rates caused due to asthma. Even the classification performance between asthmatic and healthy subjects using estimated boundaries by proposed method is found to be comparable with that of ground truth boundaries.



\section{Dataset}
For this work, data has been recorded from a total of 60 (24F, 36M) subjects, out of which 31 (12F, 19M) healthy controls and 29 (12F, 17M) patients. The subjects' age varies from 15 years to 53 years, where the average age of patients is 37.17 years, and for controls is 30.38 years.
We have recorded data in the noisy condition of the hospital, which includes noises like
people talking, fan, AC, phones ringing, etc.
A consent form has been taken from each subject before recording. All patients recording have been done in st. Johns Medical college hospital under the doctor's guidance. St. Johns medical college and hospital, Bangalore, Karnataka, India, ethics committee approved the study (Protocol number: IEC study ref no. 382/2018) on 12$^{th}$, February, 2019. All recordings have been done at the sampling rate of 44.1kHz and 16 bits using a ZOOM H6 handy recorder microphone. The microphone is kept at a distance of 3cm-5cm from the mouth while recording. While recording, the subject's nose is closed with the nose clip, breathing only through the mouth. Deep breaths of the subjects have been recorded. 5 breath sounds have been recorded on an average per subject.
Total 151 controls' and 156 patients' breath sounds have been recorded. Hence the total 307. The average duration of breath sound is 2.97 secs, and the standard deviation is 1.47 secs. The minimum and maximum duration of breaths are 1.125 secs and 11.356 secs, respectively.
\begin{figure}[h!]
	\centering
	\includegraphics[trim={3cm 0cm 0cm 0cm},scale=.2,clip=true]{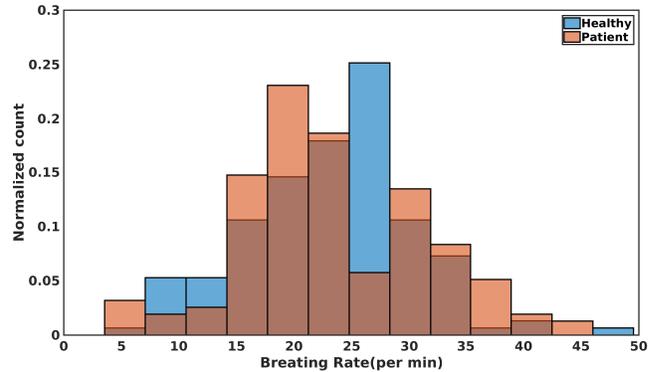}		
	\caption{Breath rate histogram between patients and healthy subjects.}
	\label{fig:Breathing_rate}
\end{figure}
Histogram of healthy subjects and asthmatic patients breating rate (given by the number of breaths per min) has been plotted in Fig. \ref{fig:Breathing_rate}. Fig. \ref{fig:Breathing_rate} shows that for asthmatic patients and healthy subjects, breathing rate is highly overlapped. For the majority of the subjects breathing rate lies between 15 to 30 breaths per minute.

\begin{figure}[h!]
	\centering
	\includegraphics[trim={4cm 8cm 4cm 4cm},scale=.4,width=\columnwidth,clip=true]{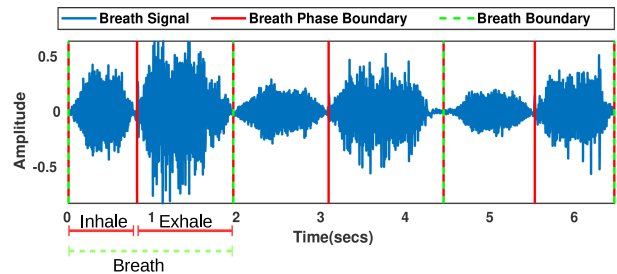}		
	\caption{Manually annotated breath sound sample file. Inhale and exhale boundaries are shown in green color and breath boundary is in red.}
	\label{fig:bre th_dur_sample}
\end{figure}

Inhale duration between subjects varies from .50 secs to 4.62 secs with an average of 1.3 secs, whereas for exhale, it varies from .65 secs to 8.09 secs with an average duration of 1.62 secs. The two annotators have annotated recorded data by inspecting the spectrogram and waveform by using Audacity \cite{mazzoni2000audacity}. An example of an annotated waveform is given in Fig. \ref{fig:bre th_dur_sample}. A noisy breath sample with 3 breaths is shown in Fig. \ref{fig:bre th_dur_sample}. Inhale and exhale boundaries are marked separately (shown in red color) for each breath, and breath boundaries are shown in green.

\section{Proposed Breath segmentation}
\begin{figure}[]
	\centering
	\includegraphics[trim={4cm 2cm 0 2cm},scale=.4,clip=true]{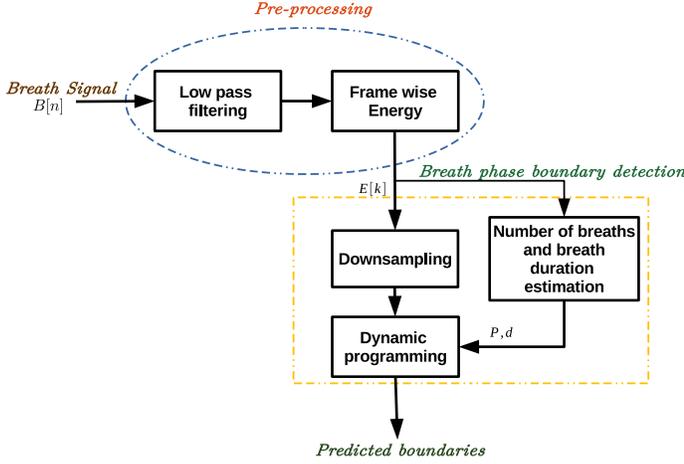}
	\caption{Block diagram of our proposed method. $B[u]$, $E[n]$, $P$ and $d$, indicates breath signal, energy signal, number of breath phases and breath phase duration, respectively. $u$ and $n$, denotes samples index and frame index.}
	\label{fig:BD}
\end{figure}

The Block diagram of our proposed method is given in Fig. \ref{fig:BD}. The proposed method has three main steps. The first is the pre-processing step, which helps remove the high-frequency noise and estimates the signal's energy. The second step involves estimating average breath duration and number of breaths in a recording and using this information to find the boundaries using dynamic programming. Each part of the block diagram is explained below in detail.
\subsection{Pre-processing}
\subsubsection{Low-pass filtering}
The energy of the breath signal (referred as \(B[u]\), where $u$ shows sample index), is present below 2kHz \cite{sarkar2015auscultation}. Therefore recorded breath sounds are low pass filtered at cut-off 2kHz. As our data has been recorded in noisy conditions, this step is necessary to remove all the noises above 2kHz to make the algorithm robust.

\subsubsection{Frame wise energy calculation}
Filtered signal has been framed with window size (\(w_s\)) and overlap (\(w_o\)). The energy of each frame has been calculated, and energy signal (\(E[n]\)) is obtained, where $n$ denotes the frame index. \(E[n]\) is used in this work to find the boundaries because it has a more stable periodic structure of breaths than raw signal, which is very noisy.
\subsection{Breath Phase boundary prediction}

\subsubsection{Breath Phase Modeling}
\label{section:breath_phase_model}
\begin{figure}[]
	\centering
	\includegraphics[trim={0cm 0cm 0cm 0cm},width=\columnwidth,clip=true]{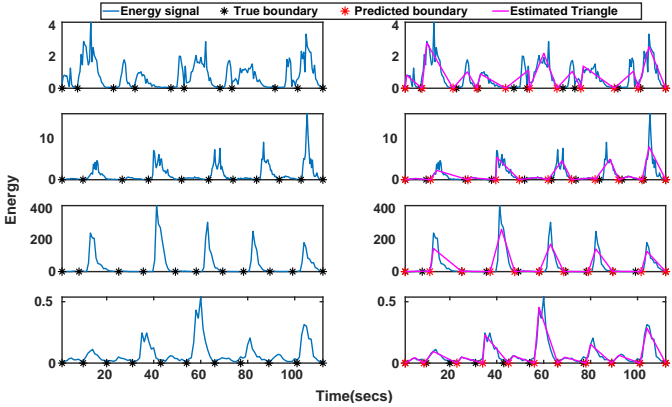}

	\caption{Samples of energy signal. The left column shows the short-time energy signal with the ground truth boundary, and the right column shows the corresponding signal with the predicted boundary using triangle fitting.}
	\label{fig:Energy_Sample}
\end{figure}
A breath has two components inhale and exhale. The energy of the breath signal taken from 4 subjects is shown in Fig. \ref{fig:Energy_Sample}. In Fig. \ref{fig:Energy_Sample} left column shows energy signal with ground truth boundaries of breath phases marked, whereas the right column shows the corresponding predicted boundaries using the proposed method in addition to ground-truth boundaries. From Fig. \ref{fig:Energy_Sample} we can see that the energy signal has some periodicity in nature. However, the amplitude of short-time energy varies a lot due to the noisy nature of recordings and breathing patterns across subjects and within a signal itself. In this work, we utilize this periodic nature of signal and energy envelop shape. In this work, we model each phase of breath by a triangle with varying duration where the endpoints of the triangle give the boundary. Hence, the short-time energy contour of a breath energy signal can be approximated by a sum of triangles, where number of triangles is equal to number of breath phases in the signal, and triangles, endpoints will give breath phase boundary. To understand how the triangle fitting has been done for each phase, consider a short-time energy contour $x[k]$ say range of $k=0,1,2,...M$, where $k$ denotes sample index. To fit a triangle to $x[k]$, between any three points $(k_1,0)$, $(k_2, x[k_2])$ and $(k_3,0)$, where \(k_1< k_2< k_3\) is given by $F[k]$ where $x[k_2]$ is indicated as $\alpha$. $F[k]$ is shown below.
\begin{figure}[ht]
	\centering
	\includegraphics[trim={5cm 6cm 4cm 4cm},scale=.4,clip=true]{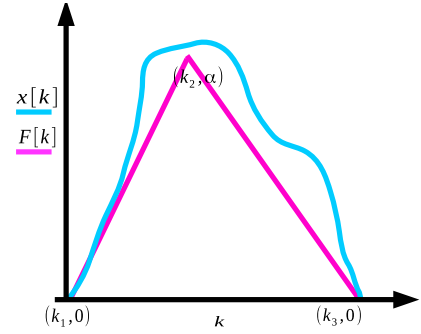}
	\caption{Demonstration of triangle fitting.}
	\label{fig:Triangle_fit}
\end{figure}
\[
F[k]=
\begin{cases}
	\frac{\alpha (k-k_1)}{k_2-k_1}, \qquad \quad \qquad k_1\leq k \leq k_2 ,\\
	\frac{\alpha (k-k_3)}{k_2-k_3}, \qquad \quad \qquad k_2 \leq k \leq k_3
\end{cases}
\]
To find the optimum $(k_2,\alpha)$ we need to minimize the following objective function.

\begin{align}
	\label{eqn:obj}
	J(k_1,k_3)=&\underset{k_2,\alpha}{\text{minimize}} \quad \sum_{k=k_1}^{k_2}(x[k]-\frac{\alpha(k-k_1)}{k_2-k_1})^2+ \\ \quad \notag
	& \quad \sum_{k=k_2+1}^{k_3}(x[k]-\frac{\alpha (k-k_3)}{k_2-k_3)})^2
\end{align}

	%
	%
By differentiating Eq. \ref{eqn:obj} with respect to $\alpha$ and equating it to zero we get the following equation in terms of $k_2$ and $\alpha$,

\begin{equation}
	\label{eqn:tr_mid}\alpha=\frac{\sum_{k=k_1}^{k_2} \frac{x[k] k}{k_2}+ \sum_{k=k_2+1}^{k_3} \frac{x[k](k-k_3)}{k_2-k_3}}{\sum_{k=k_1}^{k_2} \frac{k^2}{k_2^2}+ \sum_{k=k_2+1}^{k_3} \frac{(k-k_3)^2}{(k_2-k_3)^2}}
\end{equation}
As from Eq. \ref{eqn:tr_mid} we can see that $(k_2,\alpha)$ cannot be solved analytically, therefore $k_2$ varies from $1$ to $k_3-1$
and at each given value of $k_2$, $\alpha$ has been calculated. Hence, for given $(k_2,\alpha)$ value of Eq. \ref{eqn:obj} can be calculated. From all calculated values of the objective function in Eq. \ref{eqn:obj}, $\alpha$ and $k_2$ corresponding to the minimum value of will be the best fit.

\subsubsection{Objective function for phase segmentation in breath signal}

In this work, we have two assumptions. The first assumption is that breath starts from the first audio signal sample and ends at the last sample, which means there is no silence at the beginning and end of the signal. Secondly, the breath phases is continuous which means the endpoint of exhale is the beginning of the next inhale, and the endpoint of inhale is the beginning of next exhale. As explained in section \label{section:breath_phase_model}, triangle fitting can be a good way to find the boundaries; therefore, we can say $E[n]$ is a train of triangles where consecutive triangles share boundaries. Hence finding the boundaries is formulated in terms of minimizing the following objective function. Let's assume if we know already number of phases ($P$) in the short-time energy signal $E[n]$ where $0\leq n \leq M$, then breath phases boundaries, $n_1^*, n_2^*,...n_{P+1}^*$ can be estimated by solving Eq. \ref{eqn:dp_obj}.
\begin{align}
	\label{eqn:dp_obj}
	&n_1^*, n_2^*,...n_{P+1}^*=\underset{n_1, n_2...n_{p+1}}{\text{argmin}} \sum_{p=1}^{P}J(n_p,n_{p+1}) \\
	&\text{subject to}\quad n_1=0,n_{P+1}=M,\notag \\
	&\{n_2,n_3..n_p\}\in\{2,....M-1\}, n_{p+1}> n_p \notag
\end{align}

Eq. \ref{eqn:dp_obj} is the iterative cost function, which depends on the cost of fitting the previous breath phases; therefore, the Eq. \ref{eqn:dp_obj} can be solved using dynamic programming (DP) problem. DP \cite{bellman1954theory} is used to solve the problems, which can be broken into sub-blocks, and the solution of each sub-block is used to find the solution of the problem. Similarly, Eq. \ref{eqn:dp_obj} can be  optimized by optimizing each phase and further minimum distance path can be used to track back the optimum boundaries (Eq. \ref{eqn:dp_obj}). Steps to solve the Eq. \ref{eqn:dp_obj} by using DP is given in Algorithm \ref{DPalgo}.
\begin{algorithm}
	
	\scriptsize
	\begin{algorithmic}
		\State \textbf{Initialization}:\\
		\quad		$P$=number of phases, $d$=phase duration, $\delta$=.3\\	
		\quad		$E_d[n]$=downsampled energy signal, $0\leq n\leq M$ \\
		\quad		$O(1,n_2)=\mathop{\min}\limits_{n_2} J(1,n_2), (Eq. \ref{eqn:obj})\quad \lfloor d(1-\delta) \rfloor \leq n_2 \leq \lfloor d(1+\delta) \rfloor $\\
		\quad		$I(1,n_2)=\mathop{\arg\min}\limits_{n_2} J(1,n_2), \quad \lfloor d(1-\delta)\rfloor \leq n_2 \leq \lfloor d(1+\delta) \rfloor $		
		\State \textbf{Recursion}:\\
		\For{$k$ varies from $2$ to $P$}\\
		
		$~~~~O(k,n_{k+1})=\mathop{\min}\limits_{k\lfloor d(1-\delta)\rfloor \leq n_k\leq k\lfloor d(1+\delta)\rfloor} \left\lbrace O(k-1,n_k) + J(n_k,n_{k+1})\right\rbrace $ \newline
		$~~~~I(k,n_{k+1})=\mathop{\arg\min}\limits_{k\lfloor d(1-\delta)\rfloor \leq n_k\leq k\lfloor d(1+\delta)\rfloor} \left\lbrace O(k-1,n_k) + J(n_k,n_{k+1})\right\rbrace $ \newline				
		$~~~~\forall ~ n_{k+1} \in {(k+1)\lfloor d(1-\delta)\rfloor \leq n_{k+1}\leq (k+1) \lfloor d(1+\delta) \rfloor}$				
		\EndFor
		\State \textbf{Back tracking}:\\
		\quad$n_{P+1}^*$= $M$ 
		\For{each phase ($l$) from $P$ to 1} \newline
		$~~~~~~~~~~~~~n_l^*= I(l,n_{l+1}^*)$;\\
		\EndFor
		\State \textbf{return}: $n_1^*, n_2^*,..., n_{P+1}^*$ 
	\end{algorithmic}
	\caption{Breath phase boundary detection by solving equation Eq. \ref{eqn:dp_obj}.}
	\label{DPalgo}
\end{algorithm}

\subsubsection{Number of breaths phases (P)}
As we explained in the previous block, boundaries of $B[u]$ can be found out by using DP, but DP requires information about the average duration of the breath phase (referred to as $d$) and number of breaths phases (referred as $P$). As $E[n]$ is a nearly periodic signal, a peak in the signal's magnitude spectrum should occur at this frequency of the signal. This property of $E[n]$ has been used to find $P$. From the spectrum of $E[n]$, the peak has been picked between our data's minimum and maximum breathing rate. Maximum and minimum breath duration is 11.2 secs and 1.2 secs calculated by manually annotated data. Hence, breathing frequency is 0.833 Hz and 0.089 Hz, respectively. The frequency at which peak has been picked is used as breath frequency from which $P$ has been estimated by doubling it.
$d$ is estimated by dividing length of $E[n]$ by $P$. These estimated values of $d$ and $P$ are used as to DP.

\subsection{Reducing complexity}
To reduce the time complexity of the DP, $E[n]$ is downsampled and the search range to find the optimum $n_k$ is reduced by putting an additional constraint on $d$. Both approaches reduce the time taken by the algorithm to find the boundaries to a great extent. Both methods are described below.

\subsubsection{Downsampling}

DP finds the minimum distance path to estimate the breath phase boundaries efficiently. As $B[u]$ is recorded at a very high sampling frequency, a very high number of points to search in $E[n]$ is well for exhaustive search. Hence, it increases the time to find the boundaries. To alleviate this problem, $E[n]$ is downsampled by a factor of 10. We experimented with different downsampling factor which varies from 2 to 14 with a step size of 2, where 10 shown to be performing best.
\subsubsection{Breath Duration estimation}

Using our method to estimate $P$, value of $d$ is calculated. As prior information of $d$ would help significantly to reduce the time taken by DP to find the boundaries by limiting the exhaustive search range to fit the triangle. Another variable referred to as $\delta$ is used, which controls the search range around $d$.

\section{Experiments}
\subsection{Baseline}
To the best of our knowledge, we did not find any work that performs the segmentation of breath sounds recorded at the mouth. Hence, in this work, we used two baselines segmentation techniques used for signals other than breath sound. In addition, as our proposed method is unsupervised, both the chosen baselines are also unsupervised for a fair comparison.
First baseline (B1), proposed for phoneme segmentation by using TIMIT corpus \cite{lyons1993darpa} by Dusan et al. \cite{dusan2006relation}. B1 uses spectral transition measures as a feature to find the transition instants between phonemes \cite{dusan2006relation}. Spectral transition measure peaks at the phoneme boundaries. These transition peaks have been picked and post-processed to find the phoneme boundaries. 

Second baseline (referred as B2) proposed by Aras et al. \cite{aras2018automatic} performs segmentation of breath sound recorded at chest. This work utilizes the repeating energy pattern of a breath sound in a given signal. By using dynamic time warping, similar repeating patterns boundaries have been found in the signal. As this method only does breath segmentation and fails to do breath phase (inhale and exhale) segmentation, therefore for comparison, we used only breath sound results.

\subsection{Evaluation Metric}

To measure the performance of the algorithm, we calculated match, deletion, segment match, and insertion by comparing the locations of predicted and ground truth boundaries as described by Ghosh et al. \cite{ghosh2007speech}. If a ground truth boundary is in $\pm$ threshold of the predicted boundary, then it is called a match (M); otherwise, the ground truth boundary is considered deleted (D). Two consecutive matches are referred to as segment match (S). All the predicted boundaries which are not matched to any ground truth boundaries are known as inserted boundaries (I). All M, D, S, and I are given in percent.


%

Another evaluation metric, Overlap Rate (referred to as $OvR$), quantifies how much segment-matched breaths and their predicted counterparts overlapped. $OvR$ has been used as an evaluation metric for phoneme segmentation work \cite{paulo2004automatic}. $OvR$ is defined as the ratio of the common duration between reference and predicted boundary divided by the maximum possible duration of the breath. $OvR$ can lie between $1$ and $0$. When $OvR$ is $1$, common duration and maximum possible duration are equal, and if the common duration is equal to 0, then $OvR$ is 0.
In this work, we analyzed the mean and standard deviation of $OvR$ for breath, inhale and exhale segmentation separately.


\subsection{Inter-annotater difference}
In the current work breath, boundaries are annotated by two annotators. To find how well the inter-annotator agreement is M, I, D, and S are used. To find the mentioned metrics, one annotator marked breath boundaries is considered as reference and other annotator marked boundaries considered predicted. Mean of the M, I, D, S, and $OvR$ has been reported. From the Table \ref{tab:interannotator}, it has been observed that out of total 367 breath boundaries of 307 breaths have a very high M (97\%) with 82\% segment match. It shows that the inter-annotator agreement is good. For this work, results are reported using only one annotator boundaries as ground truth.

\begin{table}[]
	\centering
	\caption{Inter-annotator agreement interms of Mean of Match(M), Insertion(I), Deletion(D), Segment match(S) and mean and standard deviation of Overlap rate($OvR$) for segment matched breath sounds.}	
	\label{tab:interannotator}	
	\begin{tabular}{|c|c|c|c|c|c|c|}
		\hline
		\textbf{\begin{tabular}[c]{@{}c@{}}Total\\  Boundaries\end{tabular}} & \textbf{M(\%)} & \textbf{D(\%)} & \textbf{I(\%)} & \textbf{S(\%)} & \textbf{\begin{tabular}[c]{@{}c@{}}\boldmath{$OvR$}\\ mean(std)\end{tabular}} \\ \hline
		\textbf{367}                                                         & \textbf{97}    & \textbf{3}     & \textbf{3}     & \textbf{82}    & \textbf{92(11)}                                                  \\ \hline
	\end{tabular}
\end{table}

\subsection{Experimental settings}
For this work, breath sound signal ($B[u]$) has been low pass filtered at the cut-off of 2kHz by using the $6_{th}$ order Butterworth filter. To calculate energy frame-wise, $B[u]$ has been framed at $N_w=.1 sec$ with $N_w=10 ms$ shift. The energy of each frame has been calculated to compute signal $E[n]$. To decrease the time complexity of DP, $E_[n]$ is downsampled to 10 samples/sec from 100 samples/sec. $\delta=.3$ is set to reduce the complexity of DP. To estimate $P$ and $d$, FFT of the $E[n]$ has been computed with FFT points twice the length of $E[n]$. Peak has been picked between the maximum and minimum breathing frequency of 0.833 Hz and 0.089 Hz from the $E[n]$ spectrum, respectively. 

Classification setup is similar as given in \cite{yadav2018comparison}. For the classification task between asthmatic and healthy subjects, statistics of Mel-frequency Cepstral Coefficients (MFCC) namely, mean, median, mode, root mean square, variance and standard deviation have been used as features and support vector machine as the classifier. Classification accuracy has been used as the metric. MFCC matrix is computed at 20 msec window with 10 msec shift. Each train and test set have 50 and 10 subjects, respectively. Five out of six folds have 26 controls and 24 patients in the train set and 5 patients and 5 controls in the corresponding test set. The remaining one fold has 6 controls and 4 patients in the test set and 25 patients and 25 healthy subjects in the train set.

B1 has been implemented in a similar way as described by Ghosh et al. \cite{ghosh2007speech}. In B2, the search range to find the minima of energy signal has been set to $1.2$ secs to $12$ secs, and the remaining steps of the B2 algorithm are kept fixed.

\section{Results}


\subsection{Comparison of baselines and proposed method}

\begin{table}[ht!]
	\centering
	\caption{Match(M), Insertion(I), Deletion(D), Segment match(S), and mean and standard deviation (std) of Overlap rate($OvR$) for segment matched breath sounds using baselines and the proposed method.}

	\label{tab:Mean_overlap_results_all_methods}
	
	\resizebox{\columnwidth}{!}{\begin{tabular}{|c|c|c|c|c|c|c|}
			\hline
			\textbf{Methods}     & \textbf{\begin{tabular}[c]{@{}c@{}}Predicted\\ Boundaries\end{tabular}}                                            & \textbf{M(\%)} & \textbf{D(\%)} & \textbf{I(\%)} & \textbf{S(\%)} & \textbf{\begin{tabular}[c]{@{}c@{}}\boldmath{$OvR$}\\ mean(std)\end{tabular}} \\ \hline
			\textbf{B1}                                 & \textbf{1028}                       & \textbf{88}    & \textbf{12}    & \textbf{192}   & \textbf{74}    & \textbf{92(6)}                                                 \\ \hline
			\textbf{B2}                         & \textbf{428}                               & \textbf{49}    & \textbf{51}    & \textbf{68}    & \textbf{48}    & \textbf{75(31)}                                                 \\ \hline
			\textbf{\begin{tabular}[c]{@{}c@{}}Proposed\end{tabular}} & \textbf{375} &\textbf{89}    & \textbf{11}    & \textbf{13}    & \textbf{79}    & \textbf{88(13)}                                                 \\ \hline
	\end{tabular}}
	
\end{table}
\begin{figure}[ht!]
	\centering
	\includegraphics[trim={0cm .0cm 0cm 0},scale=.4,width=\columnwidth,clip=true]{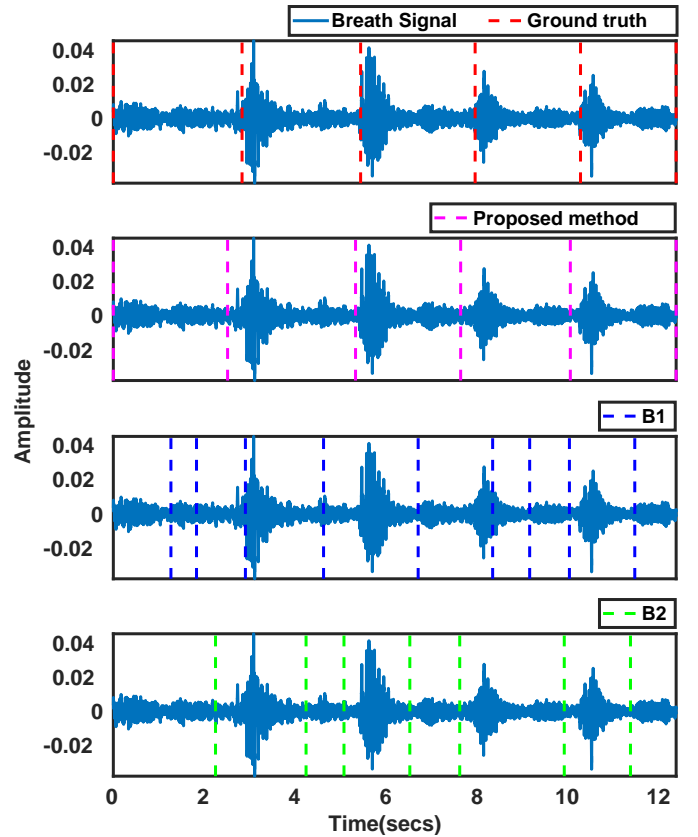}

	\caption{Example to compare the performance of the proposed method, B1 and B2.}
	\label{fig:P_B1_B2}
\end{figure}
Results for breath segmentation using the proposed method, B1 and B2, are shown in the Table. \ref{tab:Mean_overlap_results_all_methods}.
Total breaths in the data are 307, and breath boundaries are 367. Similarly, the total boundaries in each phase are 614.
From Table \ref{tab:Mean_overlap_results_all_methods} we can see that B2 performs poor among all, as it has low M ($49\%$) and S ($48\%$) compared to other two methods B1 (M ($88\%$)\& S ($74\%$)) and proposed method (M ($89\%$)\& S ($79\%$)). B1 performed comparatively well compared to B2 and the proposed method in terms of M and S but has a very high number of inserted boundaries (I), $192\%$ compared to B2. However, the B2 and proposed method has $68\%$ and $13\%$ inserted boundaries, respectively. As the TIMIT database is clean, whereas ours is a noisy database, many spurious peaks will be there while calculating spectral transition measures, which can be detected as the false peak boundaries lead to a high I measure. 
Example of the performance of the proposed method, B1 and B2, are shown in Fig. \ref{fig:P_B1_B2}. From Fig. \ref{fig:P_B1_B2} we can see that boundaries predicted by the proposed method are comparable to ground truth boundaries, whereas B1 and B2 predicted more and incorrect boundaries than the ground truth.

\subsubsection{Phase Segmentation Performance}

Results for inhale and exhale segmentation are shown in Table \ref{tab:inhale_seg.} and \ref{tab:exhale_seg.}. From the Table \ref{tab:inhale_seg.} it can be seen that B2 performed poor among all as it has minimum M and S and high D. Comparison of B1 and proposed approach shows that the proposed method performs better than the B1 even though M and S are high in B1 due to high I. Whereas I is low in the proposed method, which is $30\%$ compared to $226\%$ of B1. A similar kind of trend is observed in exhale segmentation \ref{tab:exhale_seg.}, S and M are high in 
B1 due to high I. The results show that the proposed method performs better among all the methods, but there is still a scope of improvement as S is $52\%$ and $61\%$ in inhale and exhale segmentation, respectively. Again B2 performed the worst among all methods in exhale segmentation as well. Results of breath and phase segmentation follow a very similar trend. B2 performed the worst among all three methods in all three cases, and the proposed method performed the best.
\begin{table}[!]
	\centering
	\caption{Performance of B1, B2 and proposed method for inhale segmentation.}
	\label{tab:inhale_seg.}
	\resizebox{\columnwidth}{!}{\begin{tabular}{|c|c|c|c|c|c|c|}
			\hline
			\textbf{\begin{tabular}[c]{@{}c@{}}Method \\ used\end{tabular}} & \textbf{\begin{tabular}[c]{@{}c@{}}Predicted\\ Boundaries\end{tabular}} & \textbf{M(\%)} & \textbf{D(\%)} & \textbf{I(\%)} & \textbf{S(\%)} & \textbf{\begin{tabular}[c]{@{}c@{}}\boldmath{$OvR$}\\ mean(std)\end{tabular}} \\ \hline
			\textbf{B1}                                                     & \textbf{1936}                                                           & \textbf{89}    & \textbf{11}    & \textbf{226}   & \textbf{81}    & \textbf{88(7)}                                                  \\ \hline
			\textbf{B2}                                                     & \textbf{740}                                                            & \textbf{42}    & \textbf{58}    & \textbf{79}    & \textbf{25}    & \textbf{61(37)}                                                 \\ \hline
			\textbf{Proposed}                                               & \textbf{630}                                                            & \textbf{73}    & \textbf{27}    & \textbf{30}    & \textbf{52}    & \textbf{85(15)}                                                 \\ \hline
	\end{tabular}}
\end{table}

\begin{table}[]
	\centering
	\caption{Performance of B1, B2 and proposed method for exhale segmentation.}
	\label{tab:exhale_seg.}
	\resizebox{\columnwidth}{!}{\begin{tabular}{|c|c|c|c|c|c|c|}
			\hline
			\textbf{\begin{tabular}[c]{@{}c@{}}Method \\ used\end{tabular}} & \textbf{\begin{tabular}[c]{@{}c@{}}Predicted\\ Boundaries\end{tabular}} & \textbf{M(\%)} & \textbf{D(\%)} & \textbf{I(\%)} & \textbf{S(\%)} & \textbf{\begin{tabular}[c]{@{}c@{}}\boldmath{$OvR$}\\ mean(std)\end{tabular}} \\ \hline
			\textbf{B1}                                                     & \textbf{1936}                                                           & \textbf{89}    & \textbf{11}    & \textbf{226}   & \textbf{81}    & \textbf{88(6)}                                                  \\ \hline
			\textbf{B2}                                                     & \textbf{740}                                                            & \textbf{49}    & \textbf{51}    & \textbf{72}    & \textbf{30}    & \textbf{68(35)}                                                 \\ \hline
			\textbf{Proposed}                                               & \textbf{630}                                                            & \textbf{78}    & \textbf{22}    & \textbf{25}    & \textbf{61}    & \textbf{84(13)}                                                 \\ \hline
	\end{tabular}}
\end{table}

\subsection{Performance comparison with ground truth and predicted $P$ and $d$}

Results of breath segmentation by using predicted and ground truth boundaries $P$ and $d$ are shown in Table \ref{tab:est_gt1}. From Table \ref{tab:est_gt1}, it can be seen that M and S are higher by using ground truth boundaries as compared to estimated $P$, even though mean $OvR$ is almost the same, being $88\%$ and $90\%$, respectively.
\begin{figure}[h!]
	\centering
	\includegraphics[trim={0cm 0cm 0cm 0cm},scale=.43,clip=true,width=\columnwidth]{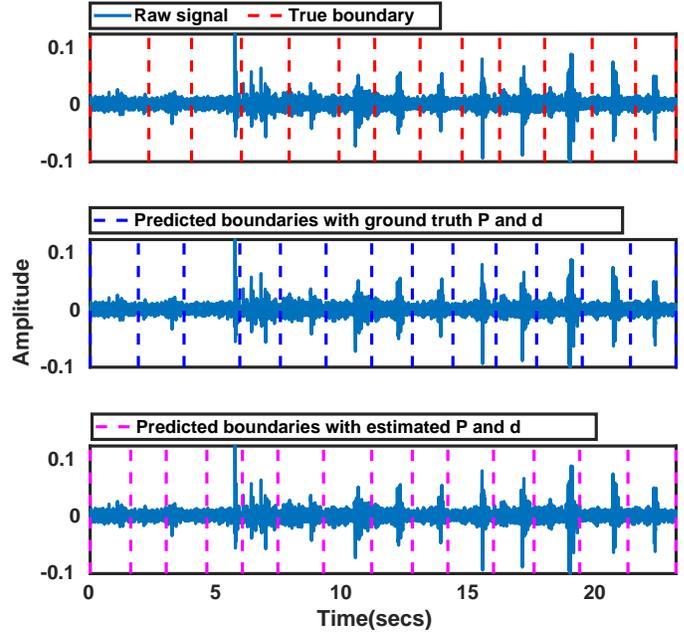}
	\caption{Predicted boundaries using predicted $P$ and $d$ performed poor compared to ground truth $P$ and $d$.}
	\label{fig:prdct_bndry_usng_estmtd_dur_Gndtrth_pass}
\end{figure}

The actual and predicted number of breaths in all 60 subjects are $307$ and $315$, respectively. Even though performance with ground truth $P$ is high, but we can see all boundary and segments matches are not close to $100\%$. Here is an example of a sample file where ground truth $P$ made poor prediction because of the first longer inhale sound than others, leading to false breath boundary prediction. Hence, all breath boundaries got displaced. In this case, we got out of 7 breath boundaries 3 to be matched, 4 deleted, and 1 segment match. This example shows that our proposed method is susceptible to the $d$ and $P$. This can be seen in another example shown in Fig. \ref{fig:prdct_bndry_usng_estmtd_dur_Gndtrth_pass}, where estimation of $P$ and $d$ is not correct, which leads to incorrect boundary prediction, whereas with correct $P$ and $d$ predictions results improved. In Fig. \ref{fig:prdct_bndry_usng_estmtd_dur_Gndtrth_pass}, predicted $P$ is 14 whereas its true value is 13, which leads to the M and D to be 7 and 7, respectively, whereas with true $P$, M, and D values are 12 and 2, respectively.\\

\begin{table}
	\centering
	\caption{Match(M), Insertion(I), Deletion(D), Segment match(S) and mean and standard deviation (in $\%$) of Overlap rate($OvR$) for segment matched breath sounds using estimated and ground truth $P$ and $d$.}
	\resizebox{\columnwidth}{!}{\begin{tabular}{|c|c|c|c|c|c|}
			\hline
			\textbf{\begin{tabular}[c]{@{}c@{}}$P$ and $d$\end{tabular}} & \textbf{M(\%)} & \textbf{D(\%)} & \textbf{I(\%)} & \textbf{S(\%)} & \textbf{\begin{tabular}[c]{@{}c@{}}\boldmath{$OvR$}\\ mean(std)\end{tabular}} \\ \hline
			\textbf{Estimated}                                                      & \textbf{89}    & \textbf{11}    & \textbf{13}    & \textbf{79}    & \textbf{88(13)}                                                 \\ \hline
			\textbf{Ground truth}                                                   & \textbf{93}    & \textbf{7}    & \textbf{7}    & \textbf{86}    & \textbf{90(7)}                                                 \\ \hline
	\end{tabular}}
	\label{tab:est_gt1}
\end{table}

\subsection{Comparison of proposed method between patients and healthy subjects}
\begin{figure}
	\centering
	\includegraphics[trim={0 4.5cm 24cm 0},width=\columnwidth, clip=true]{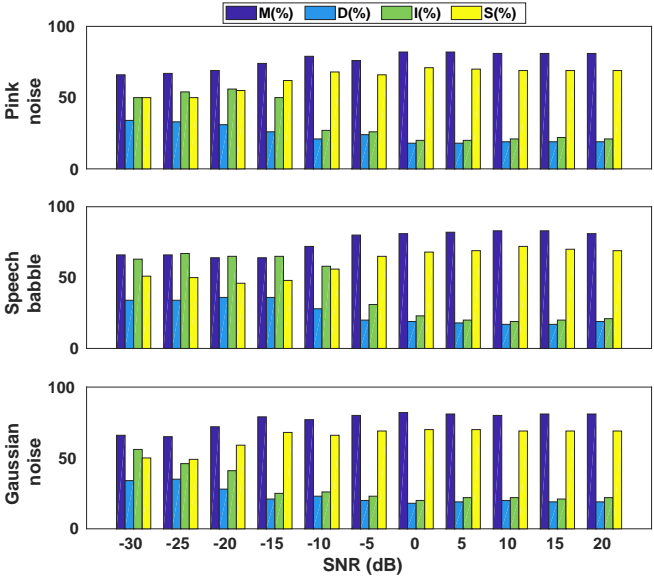}
	\caption{Comparison of evaluation metric Matching(M), Deletion(D), Insertion(I) and Segment match(S) across different noises namely, Pink noise, Speech babble and Gaussian noise added at SNR(dB) (shown as x-axis), which varies from -30 dB to 20 dB with 5dB step.}
	\label{fig:noise_comapare}
\end{figure}
As our data consists of both healthy and patients, we analyzed the proposed method performance between groups. Table \ref{tab:Estimated_bound_asthm_health.} shows the results of the proposed method in both groups. From the results, we can see that performance is better among the control group than patients as M is $98\%$ and $81\%$, respectively. In the case of patients, S is $63\%$, which is very low compared to S of control, which is $95\%$. We also observe that in the case of patients, D and I are very high compared to healthy subjects. The reason for poor performance in the case of patients is due to incorrect prediction of $P$ and hence $d$. Poor prediction of $P$ can be due to irregular breath duration within continuous breath signal for a patient compared to healthy subjects due to breathing difficulty. However, for healthy subjects, breath duration won't vary a lot.
\begin{table}
	\centering
	\caption{Match(M), Insertion(I), Deletion(D), Segment match(S) and mean and standard deviation (in $\%$) of Overlap rate($OvR$) for segment matched breath sounds in patients and healthy.}
	\label{tab:Estimated_bound_asthm_health.}	
	\resizebox{\columnwidth}{!}{\begin{tabular}{|c|c|c|c|c|c|c|}
			\hline
			\textbf{\begin{tabular}[c]{@{}c@{}}Method \\ used\end{tabular}} & \textbf{\begin{tabular}[c]{@{}c@{}}Predicted\\ Boundaries\end{tabular}} & \textbf{M(\%)} & \textbf{D(\%)} & \textbf{I(\%)} & \textbf{S(\%)} & \textbf{\begin{tabular}[c]{@{}c@{}}\boldmath{$OvR$}\\ mean(std)\end{tabular}} \\ \hline
			\textbf{Patient} & \textbf{193}&\textbf{81}    & \textbf{19}    & \textbf{24}    & \textbf{63}    & \textbf{86(17)}                                                 \\ \hline
			\textbf{Control} & \textbf{182}&\textbf{98}    & \textbf{2}     & \textbf{2}     & \textbf{95}    & \textbf{90(10)}                                                 \\ \hline
	\end{tabular}}
	
\end{table}


\subsection{Additive noise robustness of the proposed method}

As for real-time application, most of the breath sounds are recorded in a very noisy environment; therefore, to analyze the performance of the proposed method by adding three types of noises has been studied. Three kinds of noises, namely, speech babble, Gaussian, and pink noise, are added to speech signals at different SNRs. SNR varies from -30dB to 20dB with a step size of 5dB. All noisy signals are at 16kHz before adding to the signal
Performance metrics after adding all three noises to the signals are given in Fig.  \ref{fig:noise_comapare}. From Fig. \ref{fig:noise_comapare} can be observed that as SNR increases, M increases, as well as S, also increases parallelly. For example, in pink noise, M increases from $66\%$ to $81\%$ and S varies from $50\%$ to $69\%$ percent for SNR -30dB and 20dB, respectively. A similar trend is found in the case of speech babble and Gaussian noise. Another observation is with increasing SNR I decrease, which indicates that estimation of $P$ improves with positive SNR, leading to better boundary prediction. For example, in the case of speech babble, I decrease from $63\%$ to $21\%$ at SNR -30dB and 20dB, respectively. From these experiments, we concluded that the proposed algorithm performance is very sensitive to the estimation of $P$. With increasing SNR, the number of predicted breaths equals ground truth, leading to better performance.

\subsection{Asthmatic patients and healthy subjects classification}
\begin{table}[]
		\centering
	\caption{Total classification accuracy (TCA)(\%) between asthmatic patients and healthy subjects is shown for each fold using estimated boundaries and ground truth boundaries. Last column indicates the mean(std) of TCA averaged across all folds. }
	\label{tab:class_est_ground}	
	
			\resizebox{\columnwidth}{!}{
			\begin{tabular}{|c|c|c|c|c|c|c|c|}
					\hline
					\textbf{\begin{tabular}[c]{@{}c@{}}Breath\\ Boundaries\end{tabular}} & \textbf{Fold1} & \textbf{Fold2} & \textbf{Fold3} & \textbf{Fold4} & \textbf{Fold5} & \textbf{Fold6} & \textbf{\begin{tabular}[c]{@{}c@{}}Mean\\ (std)\end{tabular}} \\ \hline
					\textbf{Ground Truth}                                                & 60             & 80             & 90             & 70             & 81.81             & 66.67             & \begin{tabular}[c]{@{}c@{}}75\\ ($\pm$11)\end{tabular}       \\ \hline
					\textbf{Estimated}                                                   & 60             & 70             & 100             & 60             & 71.57             & 66.66             & \begin{tabular}[c]{@{}c@{}}72\\ ($\pm$ 15)\end{tabular}         \\ \hline
				\end{tabular}}
	\end{table}
Fold-wise classification accuracy using estimated and ground-truth boundaries have been given in the table \ref{tab:class_est_ground}. Mean TCA using estimated boundaries is 72\%($\pm$ 15\%), whereas, with ground truth boundaries, it is 75\%($\pm$ 11\%). Classification using estimated boundaries is close to the ground truth boundaries.
 Hence, the proposed method can be used for the segmentation, and segmented data can be used for the classification.

\section{Conclusion}
In this work, a breath segmentation algorithm is proposed. The proposed algorithm is used to find the boundaries of breath and its' phases inhale and exhale. The periodic nature of the signal energy is used to find the boundaries using dynamic programming. We observed that the breath signal's periodic nature could be good enough to find the breath boundaries. It has been found that the proposed method performs well in a healthy group compared to asthmatic subjects because patients have more variation in the breath phase duration than healthy subjects. Hence, for better algorithm performance robust estimation of breath phase duration and number of breath phases are required. Classification performance between asthmatic and healthy subjects is found to be comparable using estimated boundaries and ground truth boundaries. The noise robustness of the proposed method makes it useful for real-time deployment. 
\section{Conflict of interest}
None declared

\bibliographystyle{unsrt}
\bibliography{ms}	
%
%
%
%
%
%
%
%
%
%
%
%
%
%
%
%
%

\end{document}